\documentclass{svproc}
\usepackage{url}

\usepackage{color}
\usepackage{amsmath}
\usepackage{amsfonts}
\usepackage{amssymb}
\usepackage{graphicx}

\begin{document}
\mainmatter              % start of a contribution
\title{ASTRA, A Transition Density Matrix Approach to the Interaction of Attosecond Radiation with Atoms and Molecules}
\titlerunning{ASTRA}  % abbreviated title (for running head)
%                                     also used for the TOC unless
%                                     \toctitle is used
%
\author{Juan M Randazzo\inst{1,2}, 
Carlos Marante\inst{1},
Siddhartha Chattopadhyay\inst{1},
Heman Gharibnejad\inst{3,4},
Barry I Schneider\inst{3},
Jeppe Olsen\inst{5},
Luca Argenti\inst{1} \thanks{\email{luca.argenti@ucf.edu}}}
\authorrunning{Juan M Randazzo \emph{et al.}} % abbreviated author list (for running head)
%
%%%% list of authors for the TOC (use if author list has to be modified)
\tocauthor{Juan M Randazzo, 
Carlos Marante,
Siddhartha Chattopadhyay,
Heman Gharibnejad,
Barry I Schneider,
Jeppe Olsen,
Luca Argenti}
\institute{
Department of Physics and CREOL, \\ University of Central Florida, Orlando, FL 32816, USA
\and
CONICET, San Carlos de Bariloche, Río Negro 8400, Argentina
\and
National Institute of Standards and Technology,\\ Applied and Computational Mathematics Division, Gaithersburg, MD 20899, USA
\and
Computational Physics Inc. Springfield, VA 22151
\and
Department of Chemistry, Aarhus University, Aarhus 8000, Denmark, EU
}

\maketitle  

\begin{abstract}
A new formalism and computer code, ASTRA (AttoSecond TRAnsitions), has been developed to treat the interactions of short, intense radiation with molecules.  The formalism makes extensive
use of transition density matrices, computed using a state-of-the-art quantum chemistry code (LUCIA), to efficiently calculate the many-body inter-channel-coupling interactions required to simulate the highly correlated electron dynamics due to atoms and molecules exposed to attosecond laser radiation.
\end{abstract}
\section{Introduction}

Advances in laser technology have enabled the production of extreme-ultraviolet and X-ray attosecond pulses with ever increasing intensity and repetition rates~\cite{Duris2020,Saito}. As a result, statistically significant attosecond-pump attosecond-probe experiments for molecular systems of chemical relevance are now possible. This technology has opened the way to the detailed time-resolved study of photoelectron emission from valence and core orbitals, of vibronic coupling in photoemission, and of Auger cascades. Since these experiments entail the formation of highly correlated and entangled photofragments, close-coupling (CC) is an ideal approach to simulate and interpret them. CC single-ionization codes suited for small molecules in interaction with XUV pulses, such as XCHEM~\cite{Marante2017,XCHEM} and UKRmol+~\cite{Masin}, already exist. However, a common theoretical and software framework that is extensible to multiple ionization, that can represent both slow ($\sim$1~eV) and fast ($\sim$1~keV) photoelectrons and that scales favorably with molecular size is still unavailable.

Based on these requirements, we have developed an innovative approach to the CC scheme, based on the use of high-order Transition Density Matrices (TDMs) between large-scale-CI ionic states with arbitrary symmetry and multiplicity, which we have implemented in a suite of codes named ASTRA (AttoSecond TRAnsitions). These TDMs, are not typically available in standard quantum-chemistry codes.  Here, they are computed using the general string formalism~\cite{olsen_generel} of the LUCIA code~\cite{Olsen}, which enables us to evaluate inter-channel matrix elements of arbitrary many-body operators with high efficiency. Currently, ASTRA is restricted to single-ionization states but work is currently underway to extend it to double-escape channels. ASTRA employs a hybrid Gaussian-B-spline basis and hence most of the one- and two-electron integrals must be computed numerically~\cite{Masin,Gharibnejad}. In this paper, we illustrate the current capabilities of ASTRA to calculate eigen-energies and total photoionization cross sections (PICS) for several systems. First, we present results on the N$_{2}$ molecule and on the boron atom for which accurate benchmark theoretical and experimental results exist. Second, we present some calculations on the PICS in formaldehyde (H$_2$CO) as an example of an intermediate-size polyatomic molecule for which some experimental results are also available. Finally, we present the results of preliminary calculations for the PICS of magnesium porphyrin (MgH$_{12}$C$_{20}$N$_4$), a large polyatomic molecule with as many as 37 atoms.

The paper is organized as follows: in Sec.~\ref{theory} we summarize the theoretical approach to the many-electron close-coupling (CC) method, the formula needed to calculate one- and two-electron matrix elements within the CC space, expressed in terms of transition density matrices, how these quantities are used to evaluate some relevant observables such as the parameters of autoionizing states and the total photoionization cross section, and the hybrid orbital basis. Section~\ref{structure} offers a brief discussion of the workflow in ASTRA and its auxiliary codes. In Sec.~\ref{results} we show results for the nitrogen molecule, atomic boron, formaldehyde, and the magnesium-porphyrin complex. Finally, in Sec.~\ref{conclusions}, we offer our conclusions. 

\section{Theoretical Methodology}\label{theory}

\subsection{Single-ionization Close-Coupling Space}

In ASTRA, each CC channel corresponds to an antisymmetrized product of a molecular ionic state $\vert A\rangle$, with $N-1$ electrons, and a spin-orbital $\vert P\rangle$ for an additional $N$-th electron, which may either be bound to the ion or lie in the continuum. The ion and the $N$-th electron are also coupled so to give rise to a well defined spin multiplicity. In ASTRA, we consistently use the largest spatial-invariance group of the molecule that is also a subgroup of $D_{2h}$ to label all the electronic orbitals. In second quantization, we can write:
\begin{equation}\label{eq:coupledstate}
|A,p;S\Sigma\rangle=\sum_{\Sigma_A\pi}C_{S_A \Sigma_A,\frac{1}{2}\pi}^{S\Sigma} a_{p_\pi}^\dagger|A_{\Sigma_A}\rangle,
\end{equation}
where $a^\dagger_{p\pi}$ is the creator operator for an electron in the spatial orbital $p$ with spin projection $\pi\in\{\frac{1}{2},-\frac{1}{2}\}$ (spin up or spin down)~\cite{Helgaker}, and $C_{a\alpha,b\beta}^{c\gamma}$ are Clebsch-Gordan coefficients~\cite{Varshalovich}.
The correlated ionic states $|A_{\Sigma_A}\rangle$ are obtained with the LUCIA code by means of a Complete Active Space (CAS) configuration-interaction (CI) calculation. The orbitals for this CASCI can be obtained from a single-state multi-configuration self-consistent field (MCSCF) calculation using the DALTON program, or from a state-averaged MCSCF calculation using LUCIA. The state-averaged MCSCF calculations of LUCIA  allow the optimization of an ensemble energy of a set of states with identical or different spatial symmetries, multiplicities and number of electrons. 

\subsection{Operator matrix elements with TDM formalism}
Consider next the evaluation of the matrix-elements of general singlet one- and two-body operators,
\begin{eqnarray}
\hat{O}=o_{rs} a^\dagger_{r\sigma} a_{s\sigma}
\hspace{0.1\textwidth}\text{and}\hspace{0.1\textwidth} \hat{G}=\frac{1}{2}g_{pqrs}a^\dagger_{p_\theta}a^\dagger_{r_\sigma}a_{s_\sigma}a_{q_\theta},\label{eq:1and2BOp}
\end{eqnarray}
where summation over repeated indexes is assumed.  Using the commutation relations of the creation and annihilation operators, the overlap, one- and two-body CC matrix elements can be written as: 
\begin{equation}\label{eq:Overlap}
\langle A|a_{p_{\pi}} a^\dagger_{q_{\theta}}|B\rangle = s_{pq}\delta_{\pi\theta} \delta_{AB}- \rho^{BA}_{p_{\pi}q_{\theta}},
\end{equation}
 
\begin{eqnarray}\label{eq:1BOp2}
\begin{aligned}
\langle A,P | \hat{O} |B,Q\rangle = s_{pq}\delta_{\pi\theta} \langle A | \hat{O}|B\rangle+\delta_{AB}\delta_{\pi,\theta}o_{pq} \\
-\rho^{BA}_{p_\pi,r_\theta}o_{rq}-o_{ps}\rho^{BA}_{s_\pi,q_\theta}+ o_{rs} \pi^{BA}_{s_\sigma p_\pi,q_\theta r_\sigma}
\end{aligned}
\end{eqnarray}
and
\begin{eqnarray}\label{eq:tbocc}
\begin{aligned}
\langle A,P|\hat{G}|B,Q\rangle&=\langle  A | \hat{G} | B \rangle s_{pq}\delta_{\pi\theta}+\delta_{\pi\theta}[pq|rs]\rho^{BA}_{s_\rho,r_\rho}-[ps|rq]\rho^{BA}_{s_\pi,r_\theta}\\
&+[pt|rs]\pi^{BA}_{t_\pi s_\rho,r_\rho q_\theta} + [qt|rs]\pi^{BA}_{p_\pi  s_\rho,r_\rho t_\theta}-\frac{1}{2}[tu|rs]\gamma^{BA}_{u_\tau s_\rho p_\pi,t_\tau r_\rho q_\theta},
\end{aligned}
\end{eqnarray}
respectively. In eqs. (\ref{eq:Overlap}-\ref{eq:tbocc}) we have introduced the one-, two- and three-body TDMs between ionic states:
\begin{equation}\label{eqs:TDMs}
\begin{split}
\rho^{BA}_{q_{\theta},p_{\pi}}&\equiv \langle A | a^\dagger_{p_{\pi}} a_{q_{\theta}} |B\rangle,\\
\pi^{BA}_{r_{\rho}s_{\sigma},p_{\pi}q_{\theta}}&\equiv \langle A | a^\dagger_{p_{\pi}} a^\dagger_{q_{\theta}} a_{s_{\sigma}} a_{r_{\rho}}|B\rangle\\
\gamma^{BA}_{ s_\sigma t_{\tau} u_{\mu}, p_\pi,  q_\theta r_\rho}&\equiv \langle A | a^\dagger_{p_{\pi}} a^\dagger_{q_{\theta}} a^\dagger_{r_{\rho}} a_{u_{\mu}} a_{t_{\tau}}a_{s_{\sigma}}|B\rangle.
\end{split}
\end{equation}
The evaluation of CC matrix elements only require first-, second- and third-order TDMs between correlated ions, the one- and two-electron integrals between the orbitals and the matrix elements of the operators involving the ionic states.
The TDMs between ions of different spatial symmetry and spin multiplicity are the most challenging quantities to evaluate. In ASTRA, this calculation is accomplished with high efficiency using LUCIA. The density matrices are initially calculated over spin-orbitals and can then be transformed to a spin-coupled form. For single ionization, the third-order TDMs are only needed to compute the matrix elements between ionic states augmented by electrons in active orbitals in the ionic states themselves. In this case, however, there is no involvement of hybrid functions and hence all the necessary matrix elements can be directly computed within the LUCIA code. As a result, in the implementation of this method, the third-order TDMs can be and are bypassed.

The equations for general operators in a close coupling space are used to evaluate the Hamiltonian, $H$, overlap, dipole matrix elements, and complex-absorption potential between states in all the multiplicities and symmetries relevant for an atomic or molecular system of interest. These matrices are the building blocks from which it is possible to compute bound and scattering states of the system, how the system evolves under the action of external pulses, the resulting distribution of the photofragments, as well as the optical response of the system, reflected in the spectrum of the transmitted or emitted radiation.

\subsection{Observables}
In this work, we focus on the structural parameters of bound and autoionizing states for selected systems, as well as on the total single photoionization cross section of the system from its ground state. All these quantities are obtained by diagonalizing the total fixed-nuclei electronic Hamiltonian, $\tilde{H}$, in a quantization box, with the addition of a complex absorption potential, $V_{\textsc{CAP}}$, which prevents artificial reflections of the photoelectron from the box boundary,
\begin{equation}
\begin{split}
\tilde{H}&=H+V_{\textsc{CAP}}\\
H&=\sum_{A>B}\frac{Z_AZ_B}{|\vec{R}_A-\vec{R}_B|}+\sum_{i}\left[\frac{p_i^2}{2}-\sum_A\frac{Z_A}{|\vec{r}_i-\vec{R}_A|}\right]+\sum_{i>j}\frac{1}{|\vec{r}_i-\vec{r}_j|}\\
V_{\textsc{CAP}}&=-ib\sum_i\theta(r_i-R_{\textsc{CAP}})\,(r_i-R_{\textsc{CAP}})^2,\qquad  b\in\mathbb{R}_0^+
\end{split}
\end{equation}
where $Z_A$ and $\vec{R}_A$ are the charge and position of nucleus $A$, $\vec{r}_i$ and $\vec{p}_i$ are the position and momentum operator of the i-th electron, $\theta(x)$ is the Heaviside step function, $b$ is a real positive constant, and $R_{\textsc{CAP}}$ is a distance from the origin chosen a few tens of atomic units smaller than the size of the quantization box, but larger than the radius within which the ionic states have appreciable electronic density. 

Let $|\boldsymbol{\phi}\rangle = (|\phi_1\rangle,|\phi_2\rangle,\ldots)$ be the CC basis and $\mathbf{\tilde{H}}=\langle\boldsymbol{\phi}|\tilde{H}|\boldsymbol{\phi}\rangle$ be the representation of the Hamiltonian with CAPs in this basis. The diagonalization of $\mathbf{\tilde{H}}$ can be written as
\begin{equation}
    \mathbf{\tilde{H}} = \mathbf{U}_{R}\mathbf{\tilde{E}}\mathbf{U}_L^\dagger,
\end{equation}
where $\mathbf{\tilde{E}}_{ij}=\delta_{ij} \tilde{E}_i$ is the diagonal matrix of the complex eigenvalues of the projected $\tilde{H}$, whereas $\mathbf{U}_{L/R}$ are the left/right eigenvectors, normalized so that $\mathbf{U}_{L}^\dagger\mathbf{U}_{R}=\mathbf{1}$.
The complex energies thus obtained can be grouped in three types. i)~those below the first ionization threshold, which have negligible imaginary part, since they represent bound state that do not reach into the CAP region. ii)~sequences of eigenvalues that depart from each threshold and acquire rapidly large negative imaginary components, representing non-resonant continuum states. iii)~isolated complex eigenvalues $\tilde{E}_i=\bar{E}_i - i\Gamma_{i}/2$, where $\bar{E}_i=\Re e(\tilde{E}_i)$ and $\Gamma_i = -2\Im m(\tilde{E}_i)$, which are largely independent on the choice of the extinction parameter $c$, and which represent the complex energies of autoionizing states. Indeed, such states can be regarded as Siegert states~\cite{Siegert}, i.e., states that comply with outgoing boundary conditions in all ionization channels. This complex representation of the Hamiltonian allows us to write down an explicit expression for the retarded resolvent $G_0^+(E)=[E-H+i0^+]^{-1}$ in the CC basis,
\begin{equation}
    \mathbf{G}_0^+(E) = \mathbf{U}_R\frac{1}{E-\mathbf{\tilde{E}}}\mathbf{U}_L^\dagger,
\end{equation}
which we can use to compute the total photoionization cross section as a function of the photon angular frequency $\omega$, for a fixed molecular orientation, $\sigma_{\mathrm{tot}}(\omega)$
\begin{equation}\label{eq:opticaltheorem}
\begin{split}
\sigma_{\mathrm{tot}}(\omega)&=\frac{4\pi^2\omega}{c}\sum_{\alpha}|\langle\Psi_{\alpha E_g+\omega} |\hat{\epsilon}\cdot\vec{\mu}|g\rangle|^2=\\
&=\frac{4\pi^2\omega}{c}\langle g|\hat{\epsilon}\cdot\vec{\mu}\left[\int d\varepsilon\sum_\alpha |\Psi_{\alpha \varepsilon}\rangle\langle \Psi_{\alpha \varepsilon}|\delta(\varepsilon-E_g-\omega)\right]\hat{\epsilon}\cdot\vec{\mu}|g\rangle=\\
&=-\frac{4\pi\omega}{c}\Im m\left\{
\langle g|\hat{\epsilon}\cdot\vec{\mu}\left[\int d\varepsilon\sum_\alpha \frac{|\Psi_{\alpha \varepsilon}\rangle\langle \Psi_{\alpha \varepsilon}|}{E_g+\omega-\varepsilon+i0^+}\right]\hat{\epsilon}\cdot\vec{\mu}|g\rangle
\right\}=\\
&=-\frac{4\pi\omega}{c}\Im m\Big[
\langle g|\,\hat{\epsilon}\cdot\vec{\mu}\,\,G_0^+(E_g+\omega)\,\,\hat{\epsilon}\cdot\vec{\mu}\,|g\rangle
\Big],
\end{split}
\end{equation}
where $|g\rangle$ is the ground state of the target, $c\simeq 137.035$ is the speed of light, in atomic units, $\hat{\epsilon}$ is the light polarization,  $\vec{\mu}=-\sum_{i=1}^{N_e}\vec{r}_i$ is the electronic dipole moment, and $|\Psi_{\alpha E}\rangle$ are a complete set of orthonormal open single-ionization scattering states, $\langle\Psi_{\alpha E}|\Psi_{\beta E'}\rangle=\delta_{\alpha\beta}\delta(E-E')$, labelled by the channel index $\alpha$. Equation~\eqref{eq:opticaltheorem}, which is nothing other than the optical theorem applied to photoionization processes, is useful because it does not require the actual calculation of a complete set of scattering states satisfying well-defined boundary conditions. Scattering states, which are needed to compute partial differential photoelectron distributions, will be the subject of future studies.

\subsection{Orbitals and Integrals}

The selection and partitioning of orbitals play an essential role in ASTRA. We represent the ions in a basis of inactive and active self-consistent molecular orbitals, obtained as described in the previous section.
An additional set of virtual molecular orbitals are used to represent the polycentric character of the photo-electron functions. A large set of hybrid functions, which are an admixture of polycentric Gaussian and spherical B-spline functions, complete the residual set of accessible single-particle configurations from the origin to a flexible boundary $R_0$, which is typically as large as 50 Bohr radii, but it can be larger without compromising the convergence of the calculations. Finally, a set of spherical B-splines is used to span the outer region.

It is convenient to divide the whole orbital space into an internal set, whose orbitals are polycentric in character and spatially cover the molecular region, and an external set, spherical in character, which does not overlap with the ion and can extend to distances as large as several hundred atomic units. This distinction enables us to design efficient algorithms, and to reduce the size of the one- and two-electron integrals in the calculations, which are currently calculated using the public GBTO library, developed as part of the UKRmol+ suite~\cite{Masin}. Alternative algorithms to compute these integrals, based on generalizations of the Becke partitioning approach to continuum functions, are being tested~\cite{Gharibnejad}. 

\section{Software structure}\label{structure}

In this section, we provide a brief description on how various observables are computed using the ASTRA software and the codes it employs. 
The molecular geometry, charge, and Gaussian orbital basis set are detailed in the MOLECULE.INP file. 
ASTRA uses this file, together with the DALTON.INP configuration file, to run the DALTON program~\cite{dalton}, which computes the one-body and two-body electronic integrals over atomic orbitals, needed for the subsequent steps of the calculation. 
Next, ASTRA calls the LUCIA program three times. In the first call, LUCIA uses the atomic integrals generated by DALTON to perform an SCF or MCSCF calculation and determine optimized ionic wave functions. In a second call, LUCIA determines the one-body TDMs between the ionic wave functions. A final call to LUCIA computes higher-order TDMs between ions as well as the Hamiltonian and dipole matrix elements between ionic states augmented by any of the molecular orbitals that are active in the ions. The GBTO library is called to orthonormalize the hybrid basis set and evaluate the one- and two-electron integrals~\cite{Masin}. 
Subsequently, ASTRA evaluates the matrix elements of any operator required (typically, the Hamiltonian, the overlap, and the electric dipole) in the close-coupling basis generated by augmenting the ions with an electron in either an active molecular orbital, a hybrid orbital, or an external B-spline spherical function, using the formulas listed in Eqs. (\ref{eq:Overlap}-\ref{eq:tbocc}).
From the close-coupling matrix elements of these fundamental operators, ASTRA computes the field-free bound states and the Siegert states~\cite{Siegert}, as well as the total photoionization cross section of the system, to the lowest order of perturbation theory, or its transient absorption spectrum, by solving the time-dependent Schr\"{o}dinger equation in the presence of arbitrary external fields.
The implementation of a scattering-states solver, is under way.

\section{Results}\label{results}

In this section we illustrate some of the capabilities of ASTRA by applying it to a few relevant systems: the nitrogen molecule, the boron atom, and the formaldehyde molecule, for which several data from the literature are available, as well as Mg-porphyrin, a large biologically relevant organo-metallic complex.

\subsection{Nitrogen}

The N$_2$ molecule is an attractive benchmark for the fixed-nuclei calculation of the total photoionization cross section since, due to the relatively large mass of its nuclei, the energy separation between different vibrational states is smaller than the Auger width of some autoionizing states that dominate the spectrum. Here, we focus on the energy region within 20~eV from the ground state, where the states are well approximated by single excitations. For this work, we use a cc-pVTZ basis for the ionic orbitals, a quantization box with maximum radius $R_{\mathrm{box}}=400$~a.u., and a maximum orbital angular momentum for both the internal and the external B-spline basis $\ell_{\mathrm{max}}=3$. The internuclear distance is chosen to coincide with the equilibrium distance determined experimentally, $R_{\mathrm{N-N}}=1.098$\,\AA~\cite{Huber_book}.

\begin{figure}
\centering
\includegraphics[width=0.475\textwidth]{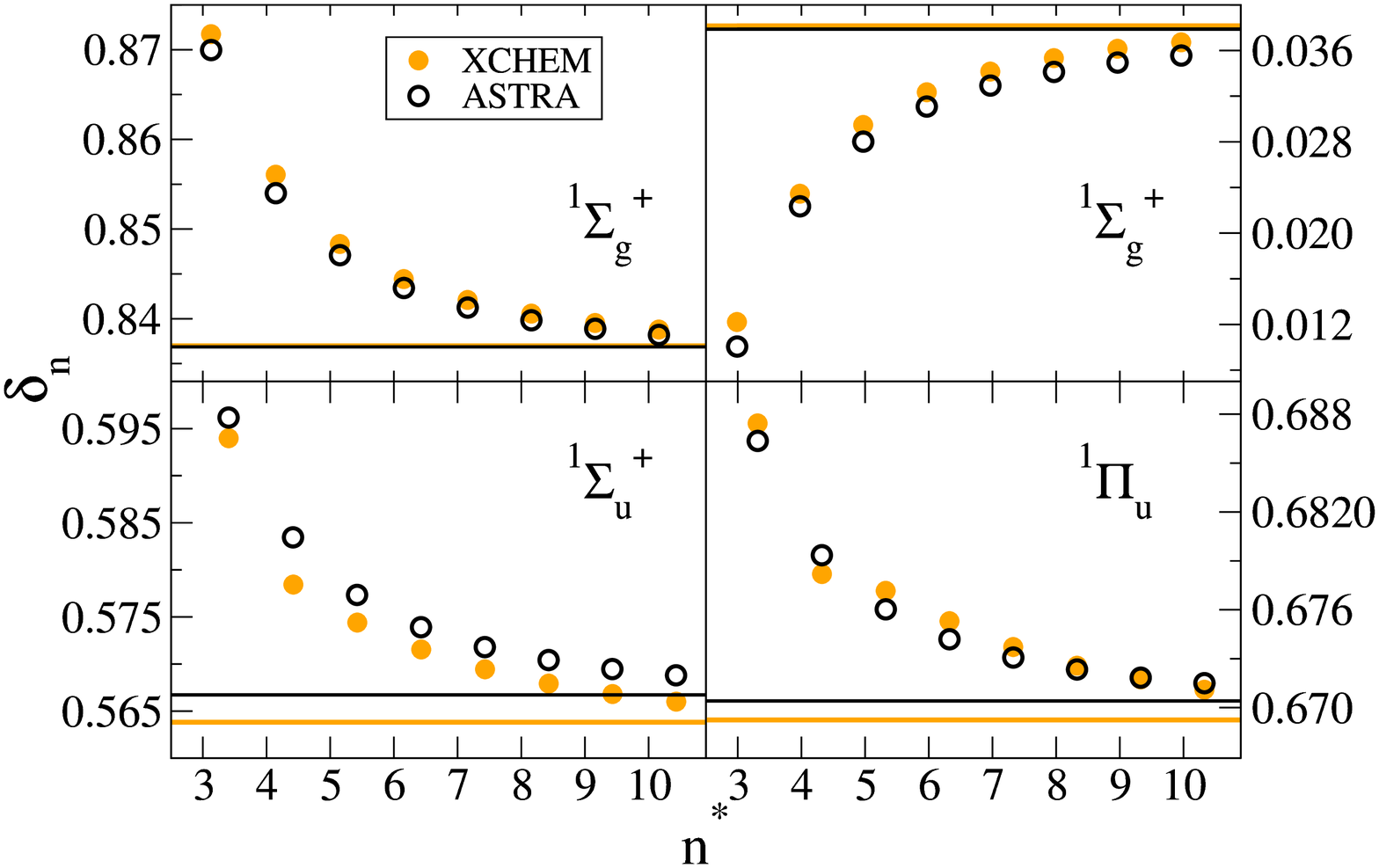}\hspace{0.005\textwidth}\includegraphics[width=0.475\textwidth]{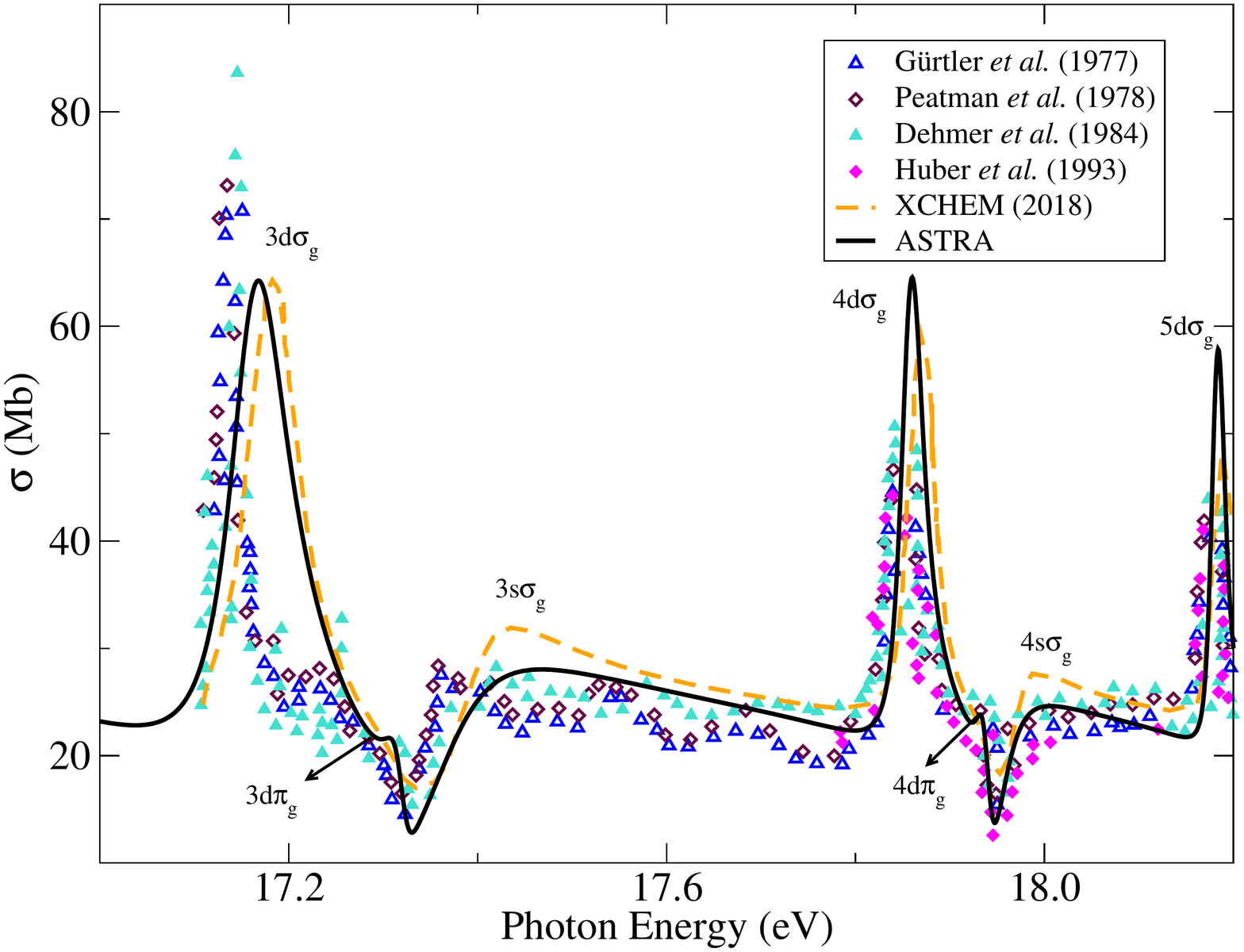}
\caption{\label{fig:N2} Left: Comparison of the quantum defect computed with ASTRA (hollow black circle), which corresponds to one of the $^1\mathrm{\Sigma^+_g}$ Rydberg series converging to the first ionization threshold of N$_2$, with the XCHEM benchmark (solid orange circle). The horizontal lines indicate the asymptotic value the series converge to. Right: Total photoionization cross section of N$_2$, in the energy region comprising the first few resonances converging to the third ionization threshold. The ASTRA cross section (solid black line) is compared with XCHEM (orange dashed line) \cite{XCHEM} and four measured spectra \cite{GURTLER1977245,Peatman1978,Dehmer1984,Huber1993}.}
\end{figure}
In the left panel of Figure~\ref{fig:N2}, we compare the quantum defect of the $^1\mathrm{\Sigma^+_g}$ bound states, $\delta_{n}=n-[2(E_{\mathrm{th}}-E_n)]^{-1/2}$, where $E_{\mathrm{th}}$ is the first ionization threshold, computed with ASTRA and with XCHEM~\cite{XCHEM}, finding an excellent agreement. In the right panel, we compare the XCHEM and ASTRA prediction for the total photoionization cross section between the second and third single-ionization thresholds with four different sets of experimental measurements. Overall, ASTRA is in good agreement with XCHEM and its predictions are marginally closer to the experimental results. ASTRA CC approach captures well not only the position and width of the main resonances and the absolute value of the background signal, but also the interference profiles for overlapping resonances. These results suggest that, for such a simple molecular system dominated by single excitations, the implementation and performance of ASTRA measure up with those of state-of-the-art codes.

\subsection{Boron}

Boron is an interesting atomic benchmark for ASTRA because it has three active electrons in an open valence shell, thus requiring ions in the CC space with both singlet and triplet multiplicities. This circumstance allows us to test parts of the code that are not active in the case of systems in a global singlet state, such as those examined in the previous subsection. We use two different close-coupling codes as benchmarks, which were developed independently from each other and from ASTRA: 1) the NewStock atomic photoionization code~\cite{NewStock}, and 2) a special-purpose high-precision three-active-electron code (TAEC)~\cite{ArgentiMoccia}. Accurate experimental values for bound states energies are also available. In this case, we use an aug-cc-pVTQZ Gaussian basis, $\ell_{\mathrm{max}}=3$, and $R_{\mathrm{box}}=300$\,a.u.

Table~\ref{tab:boron_en} compares the excitation energies of the first few bound states of boron, computed with ASTRA, NewStock, and TAEC, and the corresponding NIST experimental values. The CC space used in ASTRA and NewStock comprises twelve parent ions, whereas the TAEC calculations are conducted using a significantly larger CC expansion (twenty parent ions) and are limited to doublet states. The excitation energies are in the right order for the three calculations and they do not deviate from each other by more than 2\%.
\begin{table}[hbtp!]
\caption{\label{tab:boron_en}Bound state energies of B I, $E-E_{2s^2 2p}$ (eV). The data from the {\tt{TAEC}} calculations are taken from~\cite{ArgentiMoccia}, whereas the experimental value are rounded from the NIST CODATA database~\cite{NIST_ASD}.}
\centering
\begin{tabular}{lllll}
\hline\hline
	Conf. &  $\tt{ASTRA}$ & \tt{NewStock} &
    \tt{TAEC} & Exp.  \\
 \hline  \vspace{-0.25cm}\\
    $2s 2p^2$  ($^4P^e$)\phantom{00000}  & 3.584\phantom{00000}   & 3.564\phantom{00000}  &  \phantom{0.}$-$     & 3.552   \\
    $2s^2 3s$  ($^2S^e$)  & 4.869   & 4.881  & 4.958\phantom{00000}   & 4.964   \\
    $2s 2p^2$  ($^2D^e$)  & 5.916   & 5.927  & 5.939   & 5.933   \\
    $2s^2 3p$  ($^2P^o$)  & 5.993   & 5.990  & 6.021   & 6.027  \\
    $2s^2 3d$  ($^2D^e$)  & 6.707   & 6.682  & 6.785   & 6.790 \\
    \hline\hline
\end{tabular}
\end{table}
The calculation with the dedicated TAEC code are clearly in much closer agreement with the experiment, with discrepancies of the order of only $5$~meV across the board. NewStock, which is not optimized for three-active-electron systems and is used in a smaller CC basis, gives results which differ from the experiment by few tens meV. Finally, the results generated by ASTRA are of comparable quality to those computed by NewStock, which is in line with the two codes using the same number of parent ions in the CC expansion. This agreement indicates that the formulas concerning ions with different multiplicity are correctly implemented.

\subsection{Polyatomic Molecules}
In this section we present preliminary results for the total photoionization cross section of formaldehyde (H$_2$CO), a small polyatomic molecule, and of Mg-porphyrin (MgH$_{12}$C$_{20}$N$_4$), a comparatively large molecule. 

\begin{figure}
\centering
\includegraphics[width=0.475\textwidth]{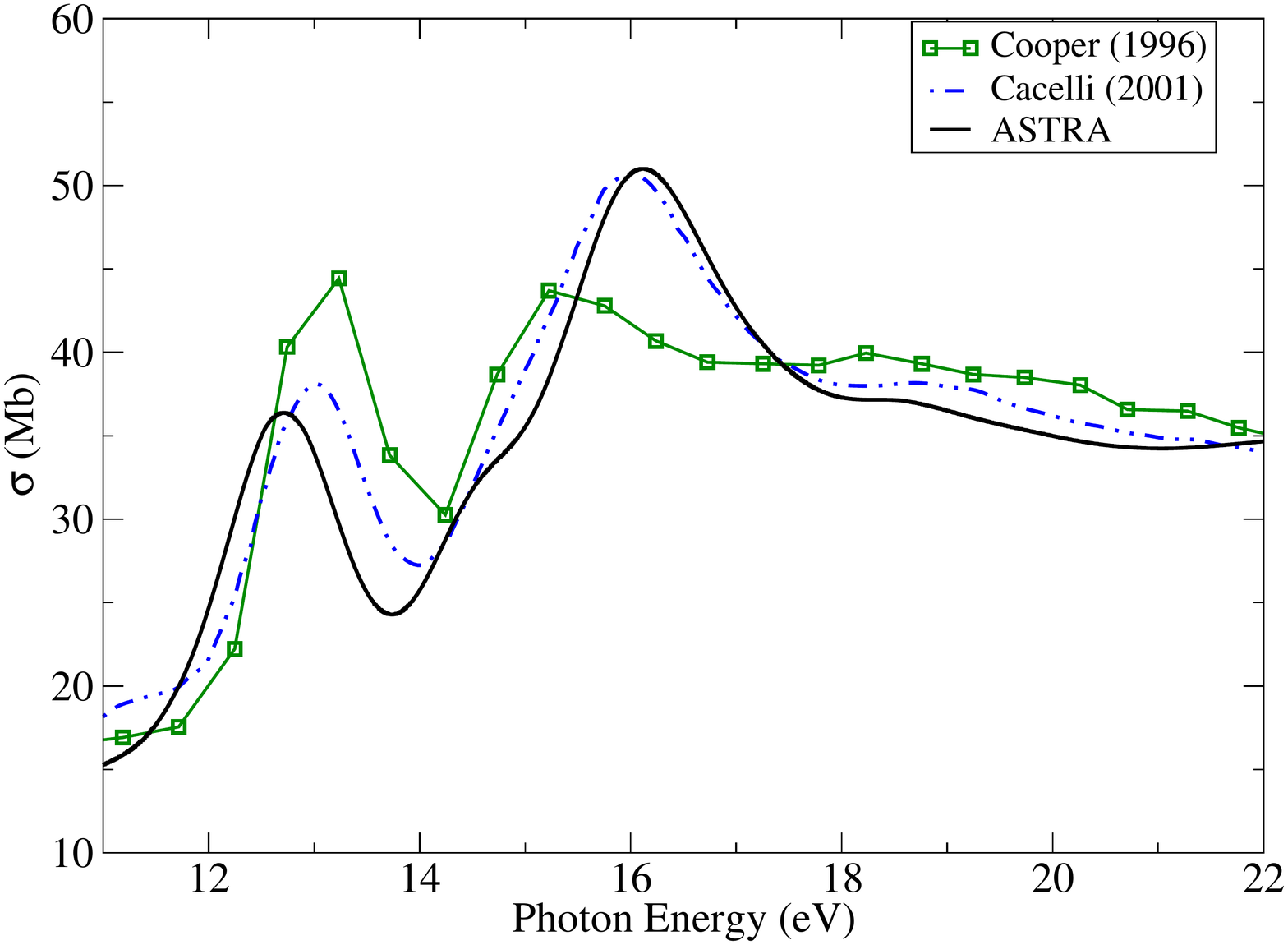}\hspace{0.005\textwidth}\includegraphics[width=0.475\textwidth]{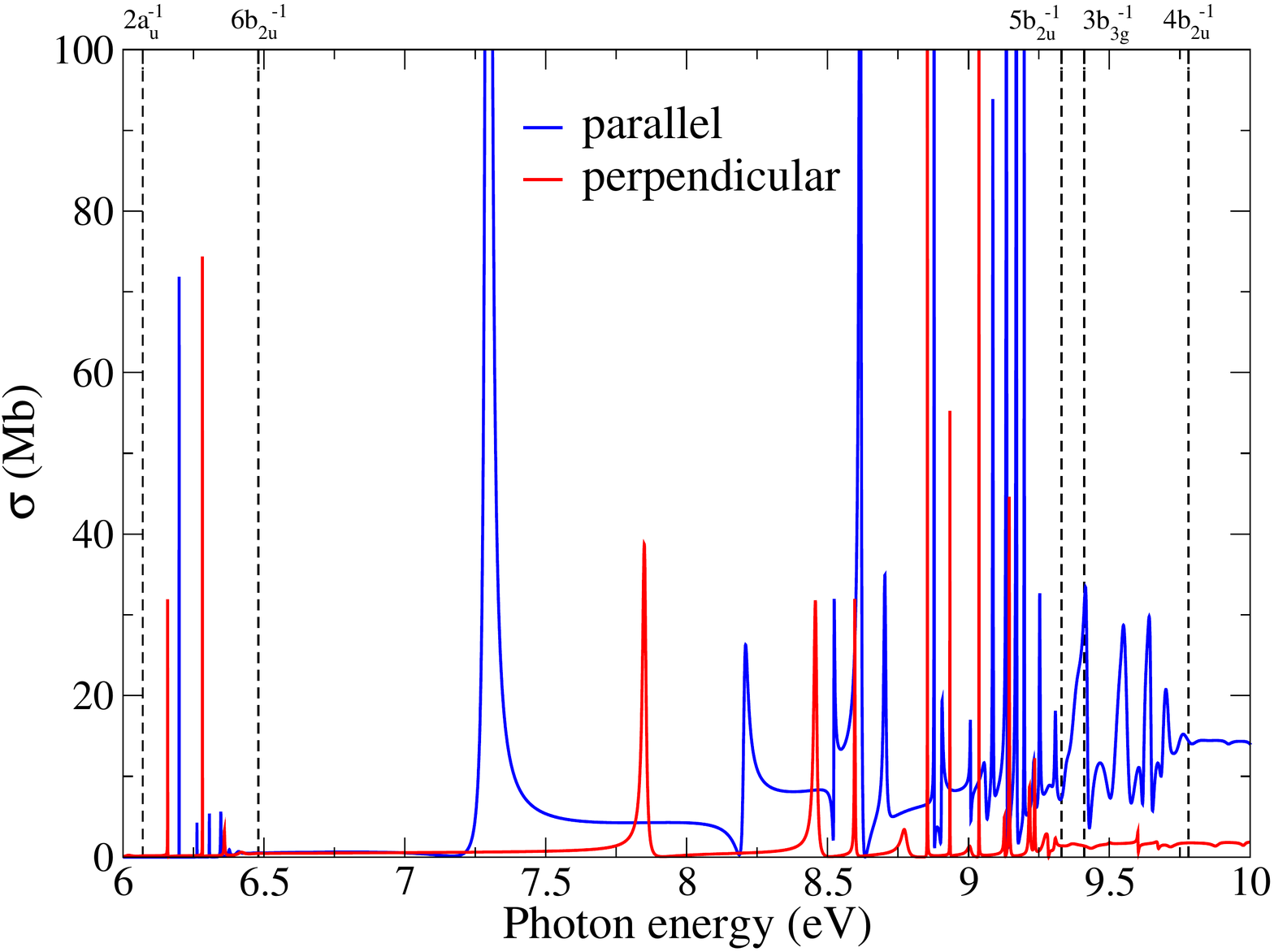}
\caption{\label{fig:formaldeidoPICS} Left: Comparison of the total PICS of formaldehyde. Both theoretical spectra, ASTRA and Cacelli \emph{et al.}~\cite{Cacelli_2001}, have been convoluted to simulate the $\sim1$ eV spectral resolution of the detector used in the experiment~\cite{Cooper1996}. Right: PICS of Mg-porphyrin, computed with ASTRA, for two different directions of the linearly polarized light, across the first six ionization thresholds.} 
\end{figure}

In the case of formaldehyde, our aim is to show that we can reproduce, with limited effort, the results of dedicated low-resolution Random-Phase-Approximation calculations~\cite{Cacelli_2001} and experimental measurements~\cite{Cooper1996}, in the photon-energy interval between 11~eV and 22~eV, as shown in the left panel of Fig.~\ref{fig:formaldeidoPICS}. 
In this case, we use a cc-pVTZ Gaussian basis, $\ell_{\mathrm{max}}=3$, and $R_{\mathrm{box}}=400$\,a.u.
The CC space used for this calculation in ASTRA includes up to the first two ions with energies above the spectral region of interest, to improve the convergence of the CC expansion. As the left panel of Fig.~\ref{fig:formaldeidoPICS} shows, the agreement between ASTRA calculations and the experiment is good and comparable to the one found by Cacelli \emph{et al.}~\cite{Cacelli_2001}. These results confirm that ASTRA is capable of giving reliable predictions even for non-linear molecules. 

Finally, our aim for Mg-porphyrin is not to demonstrate a good comparison with theoretical photoionization benchmarks (which, to the best of our knowledge, are not available) but to show that ASTRA completes all the steps needed to produce a ionization spectrum in a minimal CC basis. For the present calculation, we restrict the ions to single-determinant doublet states obtained from the Hartree-Fock state of the neutral molecule, computed in a 6-31G Gaussian basis, by removing an electron from any of the eight highest occupied molecular orbitals (i.e., from HOMO-7 to HOMO). The CC space generated from these ions, therefore, is a restricted close-coupling single-excitation (CIS) basis. In this case, $\ell_{\mathrm{max}}=1$, and $R_{\mathrm{box}}=200$\,a.u. The right panel in Fig.~\ref{fig:formaldeidoPICS} shows the contributions to the total photoionization cross section for a randomly oriented Mg-porphyrin molecule due to the light polarized either parallel or perpendicular to the $C_4$ molecular symmetry axis, across five consecutive thresholds. The plot shows several complex resonant profiles. While a converged calculation will require further work to properly account for correlation in this complex molecule, it is already clear that this system will exhibit a rich electronic dynamics that ASTRA is well positioned to analyze.

\section{Conclusions and Perspectives}\label{conclusions}

We have presented, and implemented in the ASTRA suite of codes, a new close-coupling approach for the ionization of polyatomic molecules that makes extensive use of the efficient transition density matrix formalism (TDMs), and of hybrid Gaussian/numerical basis to accurately reproduce the electronic continuum. The TDMs are obtained from LUCIA, a state-of-the-art quantum-chemistry code for large-scale CI calculations, with excellent scaling properties with respect to the size of the ionic CI expansions. ASTRA implements exact expressions for the matrix elements between arbitrary CC states, including both direct and exchange terms. We have shown that ASTRA can reproduce the results of other established theoretical methods for a few selected atomic and molecular systems for which reliable theoretical ionization cross sections exist. We have also shown that ASTRA is capable of dealing with molecules containing as many as $\sim$40 atoms.   
Work to extend ASTRA to the calculation of molecular-frame photoelectron distributions, of ionic states at the Restricted-Active-Space level, and of multiple ionization are currently under way. These extensions hinge on the development of new numerical libraries for the calculation of electronic integrals in hybrid polycentric-Gaussian/numerical bases that converge rapidly regardless of the relative position of the atoms in the system of interest~\cite{Gharibnejad}. 

\section*{Acknowledgments}
This work is supported by the DOE CAREER grant No.~DE-SC0020311.

\bibliographystyle{spphys}

\bibliography{biblio}

\end{document}